\documentclass{Interspeech}
\usepackage{times}
\usepackage{latexsym}
\usepackage{booktabs}
\usepackage{verbatim}
\usepackage[table,xcdraw]{xcolor}
\usepackage{multirow}
\usepackage[table,xcdraw]{xcolor}
\usepackage{colortbl}  
\usepackage[table,xcdraw]{xcolor}
\usepackage{booktabs}
\usepackage[T1]{fontenc}
\usepackage{etoolbox}
\makeatletter
\patchcmd{\affiliation}
  {\renewcommand{\affiliationsep}{\vskip1pt}}
  {\renewcommand{\affiliationsep}{, }}
  {}{}
\makeatother
\usepackage[utf8]{inputenc}
\usepackage{microtype}
\definecolor{lightred}{rgb}{1.0, 0.8, 0.8}
\definecolor{lightblue}{rgb}{0.8, 0.9, 1.0}
\definecolor{lightgreen}{rgb}{0.8, 1.0, 0.8}
\definecolor{lightyellow}{rgb}{1.0, 1.0, 0.8}
\definecolor{lightpurple}{rgb}{0.9, 0.8, 1.0}
\definecolor{lightorange}{rgb}{1.0, 0.9, 0.8}
\usepackage{inconsolata}
\usepackage{adjustbox}
\usepackage{subcaption, url, float, etoolbox, adjustbox, pgf,booktabs, verbatim}
\usepackage{xcolor}
\usepackage{amsmath}
\usepackage[margin=1in]{geometry}
\usepackage{amssymb}
\usepackage{tcolorbox}

\definecolor{matteyellow5}{RGB}{255,250,240}   % Very Light Matte Yellow
\definecolor{matteyellow10}{RGB}{255,245,225}  % Slightly Darker Very Light Yellow
\definecolor{matteyellow15}{RGB}{255,240,210}  % Light Matte Yellow
\definecolor{matteyellow20}{RGB}{255,235,195}  % Soft Yellow Transition
\definecolor{matteyellow25}{RGB}{255,230,180}  % Soft Matte Yellow
\definecolor{matteyellow30}{RGB}{255,225,165}  % Balanced Matte Yellow
\definecolor{matteyellow35}{RGB}{255,220,150}  % Muted Matte Yellow
\definecolor{matteyellow40}{RGB}{255,215,135}  % Gentle Transition Yellow
\definecolor{matteyellow45}{RGB}{255,210,120}  % Gentle Matte Yellow
\definecolor{matteyellow50}{RGB}{255,205,105}  % Softened Medium Yellow
\definecolor{matteyellow55}{RGB}{255,200,90}   % Medium Matte Yellow
\definecolor{matteyellow60}{RGB}{255,195,75}   % Deep Matte Yellow
\definecolor{matteyellow65}{RGB}{255,190,60}   % Rich Matte Yellow

\title{\texttt{\textbf{SNIFR}} : Boosting Fine-Grained Child Harmful Content Detection Through Audio-Visual Alignment with Cascaded Cross-Transformer}
\author[affiliation={1}]{Orchid Chetia}{Phukan}
\author[affiliation={1,2}]{Mohd Mujtaba}{Akhtar*}
\author[affiliation={1,3}]{Girish*}{}
\author[affiliation={4}]{Swarup Ranjan}{Behera}
\author[affiliation={1}]{Abu Osama}{Siddiqui}
\author[affiliation={1}]{Sarthak}{Jain}
\author[affiliation={4}]{Priyabrata}{Mallick}
\author[affiliation={4}]{Jaya Sai Kiran}{Patibandla}
\author[affiliation={5}]{Pailla}{Balakrishna Reddy}
\author[affiliation={1}]{Arun Balaji}{Buduru}
\author[affiliation={6,7}]{Rajesh}{Sharma}

\affiliation{}{IIIT-Delhi}{India}
\affiliation{}{V.B.S.P.U}{India}
\affiliation{}{UPES}{India}
\affiliation{}{Independent Researcher}{India}
\affiliation{}{Reliance AI}{India}
\affiliation{}{University of Tartu}{Estonia}
\affiliation{}{Plaksha University}{India}
\email{\textcolor{blue}{\texttt{Correspondence:}} orchidp@iiitd.ac.in}
\keywords{Child Unsafe Content, Multimodal Learning, Cross-Transformer}

\usepackage{comment}

\begin{document}
\maketitle
\maketitle
\begingroup
  % switch to symbol‐style so \footnotetext uses “*” for the first footnote
  \renewcommand{\thefootnote}{\fnsymbol{footnote}}
  \setcounter{footnote}{0}
   \footnotetext{* Equal contribution as second authors}
\endgroup
\begin{abstract}
 %In this paper, we introduce a novel framework for fine-grained detection of content considered unsafe for children. The surge in video-sharing platforms has led to a substantial increase in child viewership, necessitating systems capable of identifying unsafe content, such as violent, sexual, or explicit scenes. Further, sometimes malicious users may strategically embed such content within videos, using only a few frames to evade moderation. Previous research has led to sufficient development in this direction of fine-grained detection of such content. However, most studies have overlooked the potential of audio cues and consider mainly the visual modality for detection. In our work, we propose a multimodal approach, combining audio and visual modalities. Our experiments demonstrate that integrating audio-visual cues leads to a significant improvement in detecting child-unsafe content compared to using visual cues alone.  \newline\textbf{Main Reference for writing the paper: MM-NodeFormer: Node Transformer Multimodal Fusion for Emotion Recognition in Conversation (INTERSPEECH 24)} \newline

\noindent As video-sharing platforms have grown over the past decade, child viewership has surged, increasing the need for precise detection of harmful content like violence or explicit scenes. Malicious users exploit moderation systems by embedding unsafe content in minimal frames to evade detection. While prior research has focused on visual cues and advanced such fine-grained detection, audio features remain underexplored. In this study, we embed audio cues with visual for fine-grained child harmful content detection and introduce \texttt{\textbf{SNIFR}}, a novel framework for effective alignment. \texttt{\textbf{SNIFR}} employs a transformer encoder for intra-modality interaction, followed by a cascaded cross-transformer for inter-modality alignment. Our approach achieves superior performance over unimodal and baseline fusion methods, setting a new state-of-the-art.

\end{abstract}

\section{Introduction}
\textbf{\textcolor{red}{Warning: The following study includes visualizations of sensitive content. Readers are advised to proceed with discretion.}} \par
\begin{figure}[hbt!]
    \centering
    \includegraphics[width=0.45\textwidth]{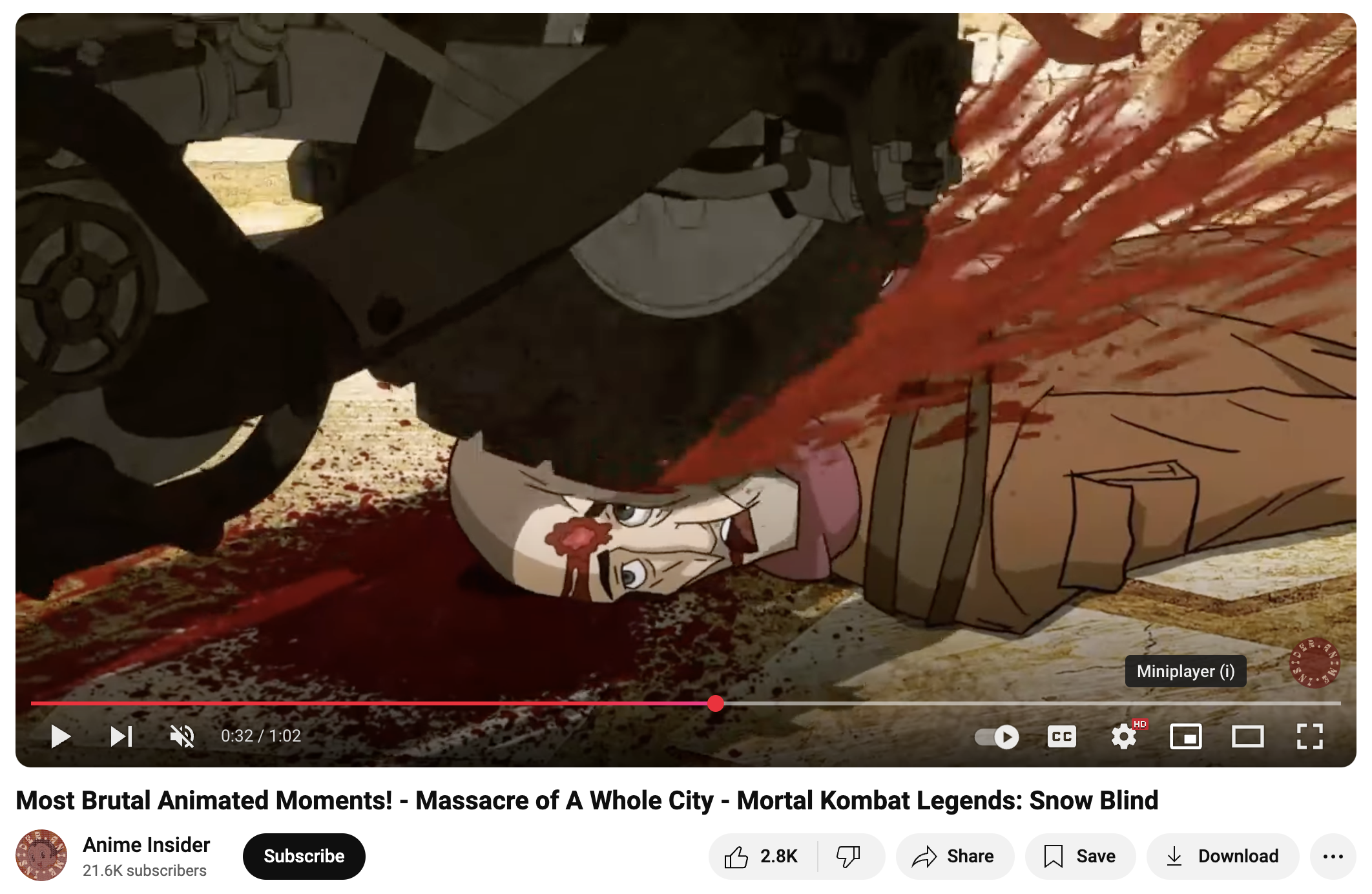} % Adjust width as needed
    \caption{Violent content in animated media}
    \label{violentimg}
\end{figure}

\noindent In today’s digital landscape, children are increasingly exposed to a vast array of video content, some of which poses significant risks to their mental and emotional well-being. Harmful content - ranging from violent scenes (Figure \ref{violentimg}) to explicit material - can have profound effects on young viewers, leading to heightened aggression, anxiety, and confusion. With platforms like YouTube, TikTok, and Netflix becoming a staple of children’s daily media consumption, often without supervision, the likelihood of encountering such content has risen dramatically. These challenges underscore the urgent need for advanced detection systems capable of reliably identifying and mitigating exposure to child-harmful content (CHC). As a remedy, researchers have explored various methods for detecting such child-harmful content (CHC) \cite{alshamrani2020hiding, aldahoul2021evaluation, chuttur2022multi, gkolemi2022youtubers, ramesh2022beach, aggarwal2023protecting, alqahtani2023children}. Kaushal et al. \cite{7906950} explored video-level features, user-level features, comment-level and CNN-based modeling with visual-features for detecting CHC. Papadamou et al. \cite{papadamou2020disturbed} made use of title, tags, thumbnail, and style features and so on from the video. Similarly, Binh et al. \cite{binh2022samba} leveraged video titles, thumbnails, and comments with LSTM and transformer networks for CHC detection. Further, Balat et al. \cite{balat2024tikguard} used video pre-trained models such as VideoMAE, TimesFormers, VIVIT for CHC detection from TikTok. \par

Despite significant advancements in content moderation, the need for fine-grained detection of CHC remains critical as malicious users may insert harmful content within videos, sometimes in as little as a few frames, in an effort to evade detection from content moderators. While some progress been made in addressing this challenge by previous researchers, particularly through visual content analysis \cite{singh2019kidsguard, yousaf2022deep}, critical gaps remain. For instance, Singh et al. \cite{singh2019kidsguard} introduced a large-scale dataset consisting of 109,835 video clips, sliced from full-length videos, aimed at fine-grained detection of child-harmful content. Their approach, utilizing an autoencoder to extract video representations followed by an LSTM classifier, effectively categorized content into safe, violent, sexual, and other harmful categories. Building on this work, Yousaf et al. \cite{yousaf2022deep} employed EfficientNet-B7 with a Bi-LSTM architecture to further improve detection accuracy. However, despite these advances, most methods have been primarily confined to visual cues, with limited exploration of audio signals as complementary information that could significantly enhance detection performance. \par

Audio signals - such as disturbing language, alarming sound effects, or suggestive tones - can provide crucial contextual information that can significantly augment fine-grained child harmful content detection (FGCHCD). This gap in the integration of audio and visual signals represents a major limitation in current detection frameworks. To address this gap, we explore fusing audio cues with visual modality and hypothesize the such integration will lead to improved performance in FGCHCD in comparison to visual-only approaches. To our end, we propose a novel framework, \texttt{\textbf{SNIFR}} (Cros\texttt{\textbf{S}}-Modality I\texttt{\textbf{N}}teract\texttt{\textbf{I}}on Cascaded Trans\texttt{\textbf{F}}o\texttt{\textbf{R}}mer) for effective alignment of audio and visual modalities. \texttt{\textbf{SNIFR}} employs a transformer encoder to capture intra-modality interactions where it efficiently models the dependencies within each modality followed by a cascaded cross-transformer that utilizes sequential cross-attention mechanisms for better inter-modality interaction. \texttt{\textbf{SNIFR}} achieves superior performance compared to visual-only, audio-only approaches and baseline fusion techniques. We also report improvement over previous state-of-the-art (SOTA) works for FGCHCD.

%we propose a multimodal approach, cross-modal interaction transformer that combines both audio and visual cues for fine-grained detection of child-unsafe content in videos. In this work, we introduce a novel framework designed to classify content as harmful or safe, thereby improving detection accuracy for child-directed platforms.%Since harmful content is often conveyed through both visual and auditory channels, a combined approach that leverages the strengths of both modalities is essential for a more comprehensive and accurate detection system. By fusing visual and auditory information, future systems can provide more precise, context-aware detection, thereby reducing the risks posed to young viewers.

\begin{figure*}[hbt!]
    \centering
    \begin{subfigure}[b]{0.3\textwidth}
        \includegraphics[width=\textwidth]{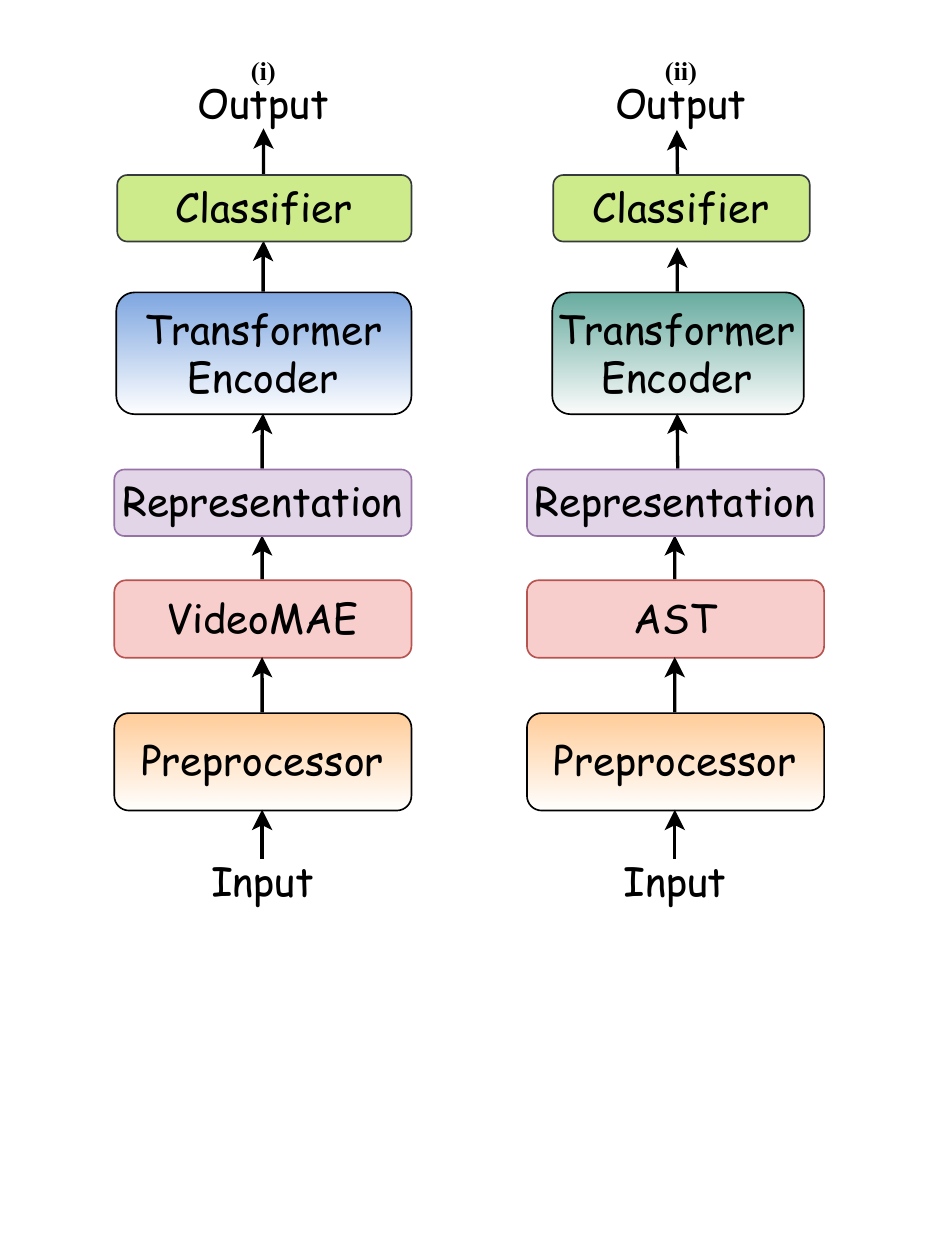}
        \caption{}
        \label{fig:image1}
    \end{subfigure}
    \hspace{0.02\textwidth} % Adjust this value to control the spacing
    \begin{subfigure}[b]{0.32\textwidth}
        \includegraphics[width=\textwidth]{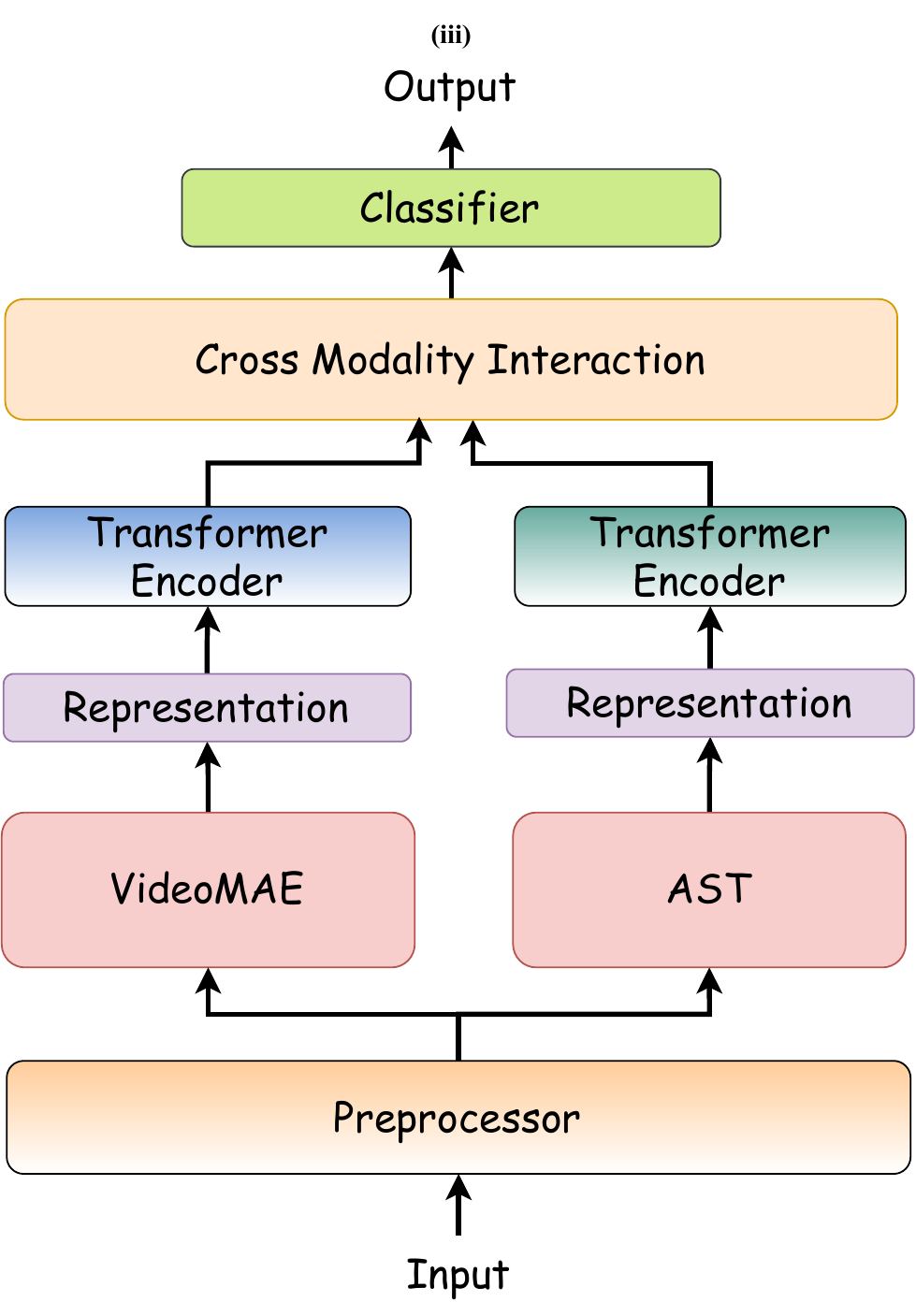}
        \caption{}
        \label{fig:image2}
    \end{subfigure}
    \hspace{0.03\textwidth} % Adjust this value to control the spacing
    \begin{subfigure}[b]{0.28\textwidth}
        \includegraphics[width=\textwidth]{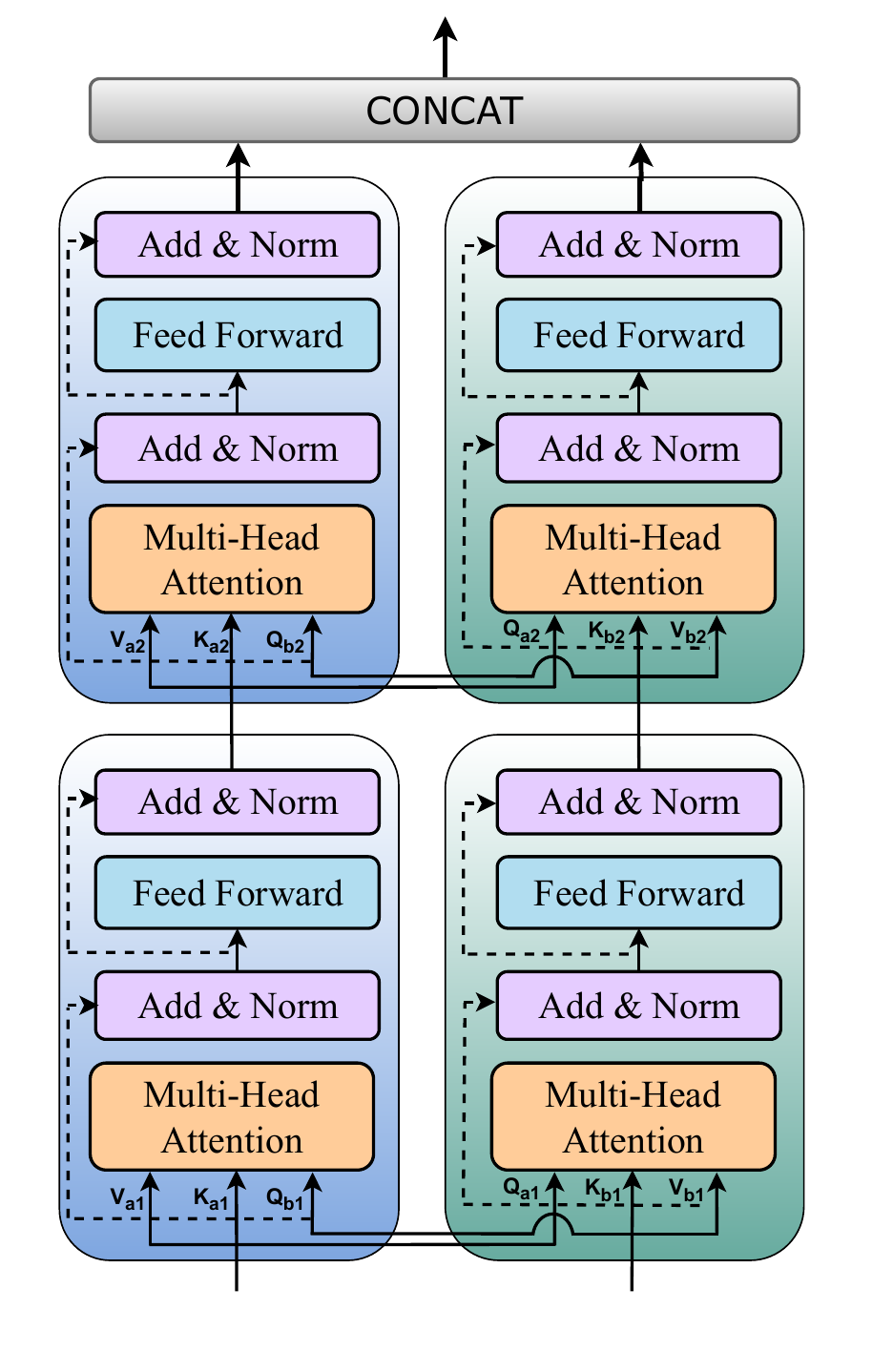}
        \caption{}
        \label{fig:image2}
    \end{subfigure}
    \caption{Modeling Architectures: Subfigure (a) show the individual unimodal modeling pipeline for video and audio, respectively; Subfigure (b) shows the proposed framework, \textbf{\texttt{SNIFR}}; Subfigure (c) provides the detailed illustration of the cross modality interaction through the cascaded cross-transformer}
    \label{architecture}
\end{figure*}

\noindent \textbf{To summarize, the main contributions of this work are as follows:}
\begin{itemize}
     \item We propose, \texttt{\textbf{SNIFR}}, a novel framework that aligns audio and visual modalities for FGCHCD. It ensures effective alignment by first using a transformer encoder to model intra-modality interactions, followed by a cascaded cross-transformer that progressively enhances inter-modality alignment through successive cross-attention mechanisms. 
    \item With \texttt{\textbf{SNIFR}}, we report topmost performance compared to unimodal video-only, audio-only as well as baseline fusion methods.  We demonstrate, for the first time to the best of our knowledge, that audio serves as a complementary modality for FGCHCD. We also report SOTA in comparison to previous SOTA approaches. 
\end{itemize}

\noindent The implementation and trained models for this work are available at: \url{https://github.com/Helix-IIIT-Delhi/SNIFR-Child_Harmful}

% The code and models developed in this work will be made publicly available following the completion of the double-blind review process.

\begin{comment}
\section{Related Work}
\textcolor{red}{O to Swwarup: Refer the cited papers in Intro and follow related papers from there. Start with child harmful content detection, from text and visual modality. Then move on to video and then move on to fine grained child harmful content detection. Then end with saying that previous studies havent considered audio, but, we do it for time.}
\end{comment}
% \section{Methodology}

% \subsection{Feature Extraction}
% \textcolor{red}{O to Swarup: Add content in AST and VideoMAE}

% \noindent \textbf{Audio Spectrogram Transformer (AST) \cite{gong21b_interspeech}}: AST is trained on AudioSet with the weights initialized from Vision Transformer. 

% \noindent \textbf{VideoMAE \cite{tong2022videomae}}: It is a representation learning model for videos pre-trained in a masked modeling manner. \newline

% \noindent We sample the video clips at 16 fps and resample the audio clips to 16KHz. We pass the audio and video clips through AST and VideoMAE, extracting embeddings of dimension size 768 via average pooling.\par

\section{Methodology}
In this section, we will discuss the feature extraction models followed by baseline models for working with unimodal features, baseline fusion techniques and finally, the proposed framework, \texttt{\textbf{SNIFR}}.
\subsection{Feature Extraction} 

\noindent \textbf{Audio Spectrogram Transformer (AST)\footnote{\url{https://huggingface.co/MIT/ast-finetuned-audioset-14-14-0.443}} \cite{gong21b_interspeech}}: It is trained on AudioSet dataset, with weights initialized from ViT, enabling it to effectively capture and encode diverse audio features. AST processes audio by first converting raw waveforms into spectrogram, which are then fed into the transformer to extract meaningful audio representations and shows SOTA behavior in diverse applications.

\noindent \textbf{VideoMAE\footnote{\url{https://huggingface.co/MCG-NJU/videomae-base}} \cite{tong2022videomae}}: It is a video representation learning model and learns to reconstruct masked portions of the input video. VideoMAE is pre-trained on large-scale video datasets, allowing it to produce generalized representations for downstream video tasks and shows SOTA performance in tasks such as video classification.

\noindent For feature extraction, we sample video clips at a frame rate of 16 frames per second and resample audio clips to a frequency of 16 kHz. These pre-processed audio and video clips are then passed through AST and VideoMAE, respectively. Each model outputs feature repersentations with a dimension of 768, achieved through average pooling across frames (for video) and time segments (for audio).

% \subsection{Model Architecture}
% \subsubsection{Audio or Visual Features}
% \noindent \textbf{Only using visual features}: The visual features were passed into a one-layer transformer followed by a two-layered feed forward network to predict the probabilities of 4 classes. 

% \noindent \textbf{Only audio features}: The audio features were passed into a one-layer transformer followed by a two-layered feed forward network to predict the probabilities of 4 classes.

\subsection{Modeling}
\subsubsection{Unimodal}
Both visual and audio modalities undergo a similar modeling pipeline (Figure \ref{fig:image1}). First, feature extraction through VideoMAE or AST and followed by vanilla transformer encoder \cite{vaswani2017attention} that models the interaction among the features. The extracted representations are then passed a classifier that contains through a dense layer with 120 neurons and finally the output layer that outputs class probabilities.

\subsubsection{Baseline Audio-Visual Fusion Methods}
We explore multiple fusion strategies for aligning audio and visual modalities for FGCHCD. In Early Concatenation (EC), the AST and VideoMAE features are passed through transformer encoder then concatenated. Similarly, in Late Concatenation (LC) AST and VideoMAE features are forwarded through transformer encoder and a dense layer of 128 neurons before concatenation. Element-wise Average (EA) fuses the transformer outputs by averaging them element-wise followed by the classifier. In Element-wise Product (EP), the modality-specific transformer outputs undergo element-wise multiplication and finally the classifier.  After concatentation in EC, and LC, it is followed by the classifier. Classifier for EC, LC, EA, EP uses the same modeling details as the unimodal modeling above. 

\subsubsection{SNIFR}
We propose a novel framework, \texttt{\textbf{SNIFR}}, for effective alignment of audio and visual modalities. The modeling architecture is illustrated in Figure \ref{architecture} 
(b). While \texttt{\textbf{SNIFR}} shares some architectural similarities with \cite{gong2022uavm}, its novelty lies in the cascaded cross-attention mechanisms, which enables more effective inter-modality interaction. The detail architectural flow of \texttt{\textbf{SNIFR}} is given as follows: First, audio and visual features are extracted using AST and VideoMAE, respectively. The intra-modality interaction is modeled using a transformer encoder that captures dependencies within each modality. Each input modality undergoes self-attention, followed by a feed-forward network (FFN) and layer normalization within the encoder. This ensures that modality-specific features are effectively learned before engaging in cross-modal interactions. For inter-modality fusion, we design a cascaded cross-transformer, which progressively aligns representations across modalities in two stages. This structured interaction ensures a gradual and effective fusion of features, allowing the model to refine relationships between modalities in a hierarchical manner. The core of the cross-transformer is the cross-attention mechanism where the query (\(\mathbf{Q}\)) is generated from one modality, while the key (\(\mathbf{K}\)) and value (\(\mathbf{V}\)) are generated from the other modality, ensuring effective cross-modal interaction. Given the outputs of the intra-modality transformer encoder, the transformations are defined as follows: $ \mathbf{Q}_A = \mathbf{Z}_A \mathbf{W}_Q^A, \quad \mathbf{K}_B = \mathbf{Z}_B \mathbf{W}_K^B, \quad \mathbf{V}_B = \mathbf{Z}_B \mathbf{W}_V^B $ and $ \mathbf{Q}_B = \mathbf{Z}_B \mathbf{W}_Q^B, \quad \mathbf{K}_A = \mathbf{Z}_A \mathbf{W}_K^A, \quad \mathbf{V}_A = \mathbf{Z}_A \mathbf{W}_V^A $. \(\mathbf{Z}_A, \mathbf{Z}_B\) are the input features for modality A (audio) and modality B (visual) after intra-modality encoding, and \(\mathbf{W}_Q, \mathbf{W}_K, \mathbf{W}_V\) are learnable weight matrices for each modality. In the first cascade, each modality attends to the other through a cross-attention mechanism, enabling the exchange of complementary features: $
    \mathbf{Z}_A^{(1)} = \text{LayerNorm} \left( \mathbf{Z}_A + \text{Attention}(\mathbf{Q}_A, \mathbf{K}_B, \mathbf{V}_B) \right) $ and 
    $\mathbf{Z}_B^{(1)} = \text{LayerNorm} \left( \mathbf{Z}_B + \text{Attention}(\mathbf{Q}_B, \mathbf{K}_A, \mathbf{V}_A) \right) $.

\noindent where the attention scores are computed as:

\begin{align}
    \text{Attention}(\mathbf{Q}_A, \mathbf{K}_B, \mathbf{V}_B) &= \text{softmax} \left( \frac{\mathbf{Q}_A \mathbf{K}_B^T}{\sqrt{d_k}} \right) \mathbf{V}_B, \\
    \text{Attention}(\mathbf{Q}_B, \mathbf{K}_A, \mathbf{V}_A) &= \text{softmax} \left( \frac{\mathbf{Q}_B \mathbf{K}_A^T}{\sqrt{d_k}} \right) \mathbf{V}_A.
\end{align}

\noindent where \( d_k \) is the dimensionality of the key vectors. After the cross-attention mechanism, the representations pass through a FFN followed by another layer normalization: $
    \mathbf{Z}_A^{(1)} = \text{LayerNorm} \left( \mathbf{Z}_A^{(1)} + \text{FFN}(\mathbf{Z}_A^{(1)}) \right) $ and 
    $\mathbf{Z}_B^{(1)} = \text{LayerNorm} \left( \mathbf{Z}_B^{(1)} + \text{FFN}(\mathbf{Z}_B^{(1)}) \right) $. The second cascade further refines this interaction by reinforcing inter-modal dependencies. The representations from the first cascade are used as input, and another round of cross-attention is applied: $ \mathbf{Z}_A^{(2)} = \text{LayerNorm} \left( \mathbf{Z}_A^{(1)} + \text{Attention}(\mathbf{Q}_A^{(1)}, \mathbf{K}_B^{(1)}, \mathbf{V}_B^{(1)}) \right) $ and 
    $ \mathbf{Z}_B^{(2)} = \text{LayerNorm} \left( \mathbf{Z}_B^{(1)} + \text{Attention}(\mathbf{Q}_B^{(1)}, \mathbf{K}_A^{(1)}, \mathbf{V}_A^{(1)}) \right) $. 

\noindent To further enhance feature refinement, another FFN is applied: $
    \mathbf{Z}_A^{(2)} = \text{LayerNorm} \left( \mathbf{Z}_A^{(2)} + \text{FFN}(\mathbf{Z}_A^{(2)}) \right) $ and $ \mathbf{Z}_B^{(2)} = \text{LayerNorm} \left( \mathbf{Z}_B^{(2)} + \text{FFN}(\mathbf{Z}_B^{(2)}) \right) $. This iterative refinement ensures that higher-order dependencies between modalities are captured, leading to a deeper integration of complementary features. After two stages of cascaded cross-transformer processing, the final fused representation is obtained by concatenating the refined outputs from both modalities: $
    \mathbf{Z}_{\text{fused}} = \text{Concat}(\mathbf{Z}_A^{(2)}, \mathbf{Z}_B^{(2)}) $. The fused representation is then fed to the classifier with the same modeling details as used in unimodal modeling above. \textbf{\texttt{SNIFR}} constitutes of 8.9M trainable parameters.

\begin{table*}[!bt]
\setlength{\tabcolsep}{8pt}
\centering
\scriptsize
\begin{tabular}{l|ccc|ccc|ccc|ccc}
\toprule
\textbf{Modality} & \multicolumn{3}{c|}{\textbf{Safe}} & \multicolumn{3}{c|}{\textbf{Sexual}} & \multicolumn{3}{c|}{\textbf{Violent}} & \multicolumn{3}{c}{\textbf{Both}} \\ 
\midrule

                  & \textbf{ACC} & \textbf{F1} & \textbf{AUC} & \textbf{ACC} & \textbf{F1} & \textbf{AUC} & \textbf{ACC} & \textbf{F1} & \textbf{AUC} & \textbf{ACC} & \textbf{F1} & \textbf{AUC} \\
\midrule
\textbf{\textit{V}}             & \cellcolor{matteyellow30}85.45 & \cellcolor{matteyellow35}89.48 & \cellcolor{matteyellow40}90.55 & \cellcolor{matteyellow40}90.70 & \cellcolor{matteyellow20}73.68 & \cellcolor{matteyellow40}95.37 & \cellcolor{matteyellow15}66.18 & \cellcolor{matteyellow10}56.96 & \cellcolor{matteyellow30}87.63 & \cellcolor{matteyellow10}64.71 & \cellcolor{matteyellow15}64.08 & \cellcolor{matteyellow35}91.49 \\
\textbf{\textit{A}}             & \cellcolor{matteyellow20}71.27 & \cellcolor{matteyellow25}80.29 & \cellcolor{matteyellow20}77.92 & \cellcolor{matteyellow30}83.33 & \cellcolor{matteyellow15}58.82 & \cellcolor{matteyellow25}82.87 & \cellcolor{matteyellow10}46.48 & \cellcolor{matteyellow5}35.11 & \cellcolor{matteyellow15}74.81 & \cellcolor{matteyellow5}50.00 & \cellcolor{matteyellow5}27.91 & \cellcolor{matteyellow20}78.58 \\
\textbf{\textit{AV (EC)}}       & \cellcolor{matteyellow35}87.19 & \cellcolor{matteyellow30}87.85 & \cellcolor{matteyellow35}88.93 & \cellcolor{matteyellow20}72.73 & \cellcolor{matteyellow20}71.91 & \cellcolor{matteyellow40}96.58 & \cellcolor{matteyellow15}58.88 & \cellcolor{matteyellow15}61.17 & \cellcolor{matteyellow30}87.01 & \cellcolor{matteyellow10}48.48 & \cellcolor{matteyellow10}40.51 & \cellcolor{matteyellow35}89.10 \\
\textbf{\textit{AV (LC)}}       & \cellcolor{matteyellow30}82.29 & \cellcolor{matteyellow30}84.40 & \cellcolor{matteyellow30}85.62 & \cellcolor{matteyellow35}85.71 & \cellcolor{matteyellow25}75.00 & \cellcolor{matteyellow35}91.06 & \cellcolor{matteyellow15}58.25 & \cellcolor{matteyellow10}56.34 & \cellcolor{matteyellow25}83.85 & \cellcolor{matteyellow10}60.00 & \cellcolor{matteyellow10}57.14 & \cellcolor{matteyellow35}92.95 \\
\textbf{\textit{AV (EA)}}       & \cellcolor{matteyellow25}79.91 & \cellcolor{matteyellow30}84.99 & \cellcolor{matteyellow30}86.67 & \cellcolor{matteyellow30}81.48 & \cellcolor{matteyellow15}60.27 & \cellcolor{matteyellow30}88.13 & \cellcolor{matteyellow10}53.42 & \cellcolor{matteyellow10}46.43 & \cellcolor{matteyellow25}85.38 & \cellcolor{matteyellow10}53.85 & \cellcolor{matteyellow10}46.67 & \cellcolor{matteyellow30}89.09 \\
\textbf{\textit{AV (EP)}}       & \cellcolor{matteyellow25}78.33 & \cellcolor{matteyellow25}81.23 & \cellcolor{matteyellow25}83.60 & \cellcolor{matteyellow20}70.59 & \cellcolor{matteyellow15}58.54 & \cellcolor{matteyellow40}95.28 & \cellcolor{matteyellow10}45.45 & \cellcolor{matteyellow5}44.55 & \cellcolor{matteyellow20}81.15 & \cellcolor{matteyellow10}51.66 & \cellcolor{matteyellow10}45.36 & \cellcolor{matteyellow30}88.29 \\
\textbf{\textit{AV (CT)}}       & \cellcolor{matteyellow30}84.65 & \cellcolor{matteyellow30}87.92 & \cellcolor{matteyellow35}90.62 & \cellcolor{matteyellow30}81.25 & \cellcolor{matteyellow20}71.23 & \cellcolor{matteyellow40}96.83 & \cellcolor{matteyellow20}66.02 & \cellcolor{matteyellow20}65.38 & \cellcolor{matteyellow35}89.83 & \cellcolor{matteyellow25}79.07 & \cellcolor{matteyellow20}64.76 & \cellcolor{matteyellow40}95.34 \\
\textbf{\textit{AV (\textit{SNIFR})}}  & \cellcolor{matteyellow40}\textbf{88.24} & \cellcolor{matteyellow40}\textbf{91.49} & \cellcolor{matteyellow40}\textbf{95.28} & \cellcolor{matteyellow40}\textbf{93.33} & \cellcolor{matteyellow30}\textbf{82.11} & \cellcolor{matteyellow40}\textbf{98.72} & \cellcolor{matteyellow35}\textbf{84.15} & \cellcolor{matteyellow35}\textbf{77.09} & \cellcolor{matteyellow40}\textbf{96.19} & \cellcolor{matteyellow35}\textbf{79.59} & \cellcolor{matteyellow30}\textbf{75.73} & \cellcolor{matteyellow40}\textbf{97.82} \\
\textbf{\textit{SOTA}~\cite{singh2019kidsguard}}  & - & - & \cellcolor{matteyellow30}88.00 & - & - & \cellcolor{matteyellow40}95.00 & - & - & \cellcolor{matteyellow35}90.00 & - & - & \cellcolor{matteyellow35}91.00 \\
\bottomrule
\end{tabular}
\caption{Evaluation results showing Accuracy (ACC) in \%, Macro-average F1-scores (F1), and AUC-scores (AUC) in \% across different classes; `Both' represents video clips containing both violent and sexual content; V: Visual, A: Audio, AV: Audio-Visual; Scores are presented in average of five-folds; AV (EC), AV (LC), AV (EA), AV (EP), AV (CT), AV (SNIFR) presents the results for audio-visual modeling with Early Concatenation, Late Concatenation, Element-wise Average, Element-wise Product, Cross-Transformer, and proposed novel framework, \textbf{\texttt{SNIFR}}}
\label{results}
\end{table*}

\section{Experiments}

\subsection{Dataset}

\begin{figure}[!bt]
    \centering
    \includegraphics[width=0.8\linewidth]{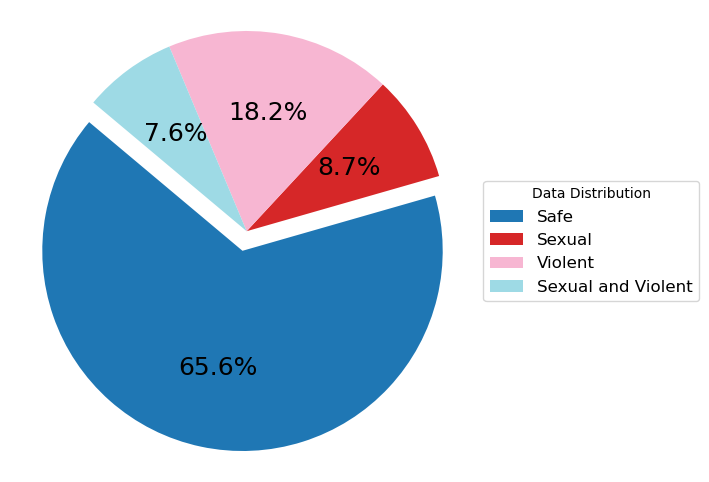}
    \caption{Percentage Distribution}
    \label{pie_plot}
\end{figure}

We conduct our experiments using the dataset presented by Singh et al.~\cite{singh2019kidsguard}, which is specifically curated for FGCHCD. We got hold of the dataset by signing a consent form given by the corresponding author. To the best of our knowledge, this is the only accessible dataset for FGCHCD with both audio and visual information present. This dataset comprises 107907 one-second video clips, divided into four categories: 70741 safe clips, 9335 containing sexual content, 19658 containing violent content and 8173 containing both sexual and violent content. The percentage distribution plot is given in Figure \ref{pie_plot}. We use FFmpeg to extract corresponding audio from each video segment.

\begin{comment}
\begin{table}[hbt!]
\centering
\caption{Evaluation results showing F1-scores in \%; 'Both' represents video clips where both violent and sexual content are present; V: Visual, A: Audio, AV: Audio-Visual}
\label{tab:scores}
\begin{tabular}{l|l|l|l|l}
\hline
\textbf{Modality} & \textbf{Safe} & \textbf{Sexual}  & \textbf{Violent} & \textbf{Both}\\ \hline
\textbf{\textit{Visual}} & 89.48 &  73.68 & 56.96 & 64.08 \\ 
\textbf{\textit{Audio}} & 80.29 & 58.82 & 35.11 & 27.91 \\ 
\textbf{\textit{Audio-Visual}} &\textbf{91.49} & \textbf{82.11}& \textbf{77.09}& \textbf{75.73} \\\hline
\end{tabular}
\end{table}
\end{comment}

% \subsection{Training Details}

% For training, we use cross-entropy loss and AdamW optimizer with a learning rate of 1e-4 and weight decay of 1e-5. The model is trained for 100 epochs.

\subsection{Training Details}
We use cross-entropy loss function, AdamW as the optimizer and setting the learning rate to 1e-4. We train the models for 25 epochs with batch size of 16. We apply a weight decay of 1e-5, dropout, and early stopping to prevent overfitting. We use 5-fold cross-validation for training and evaluation of our models where four folds are used for training and one fold for evaluation.

\subsection{Experimental Results}

\begin{figure}[!bt]
    \centering
    \subfloat[]{%
        \includegraphics[width=0.2\textwidth]{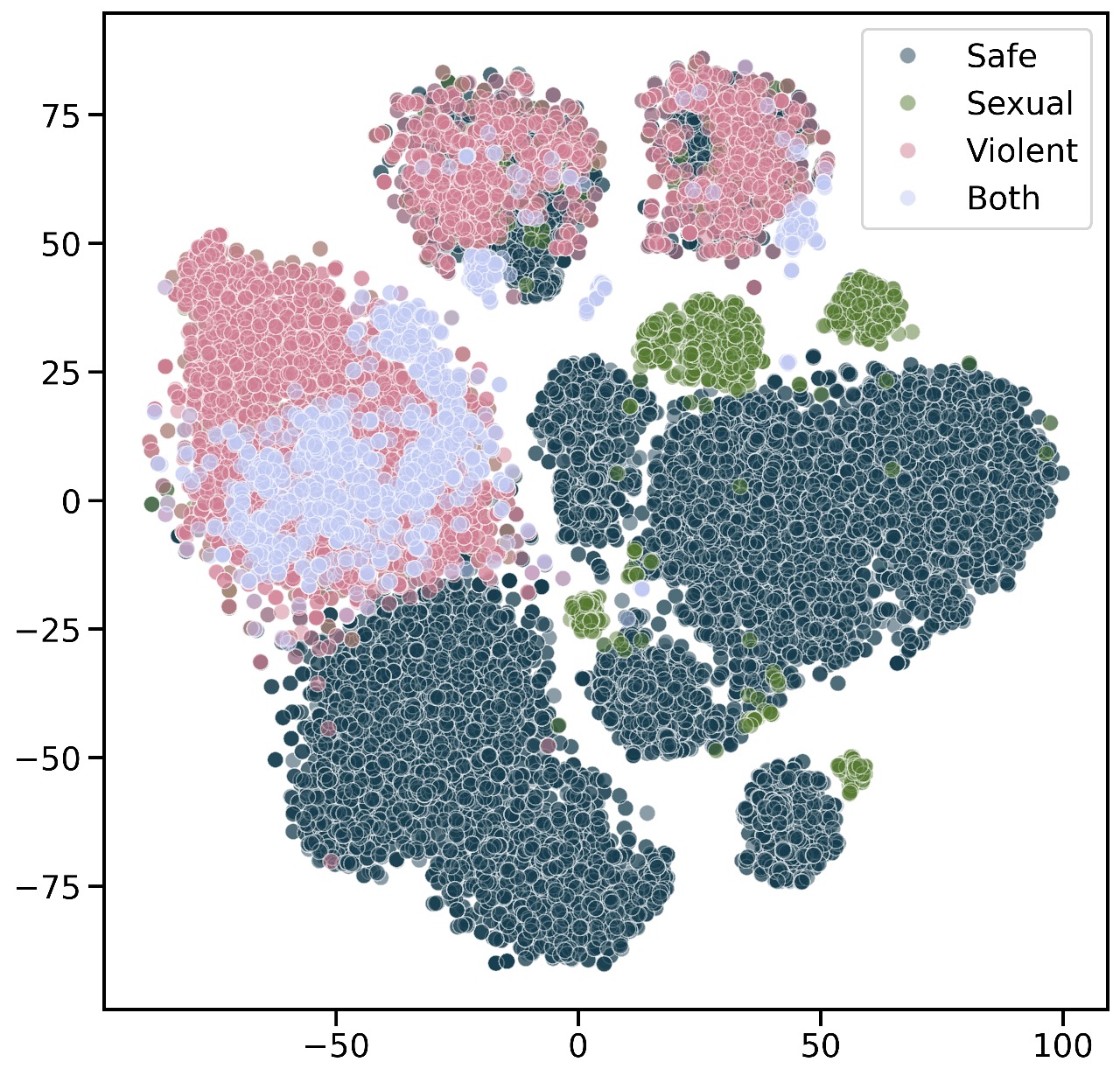}
    }
    \hspace{0.03\textwidth}
    \subfloat[]{%
        \includegraphics[width=0.2\textwidth]{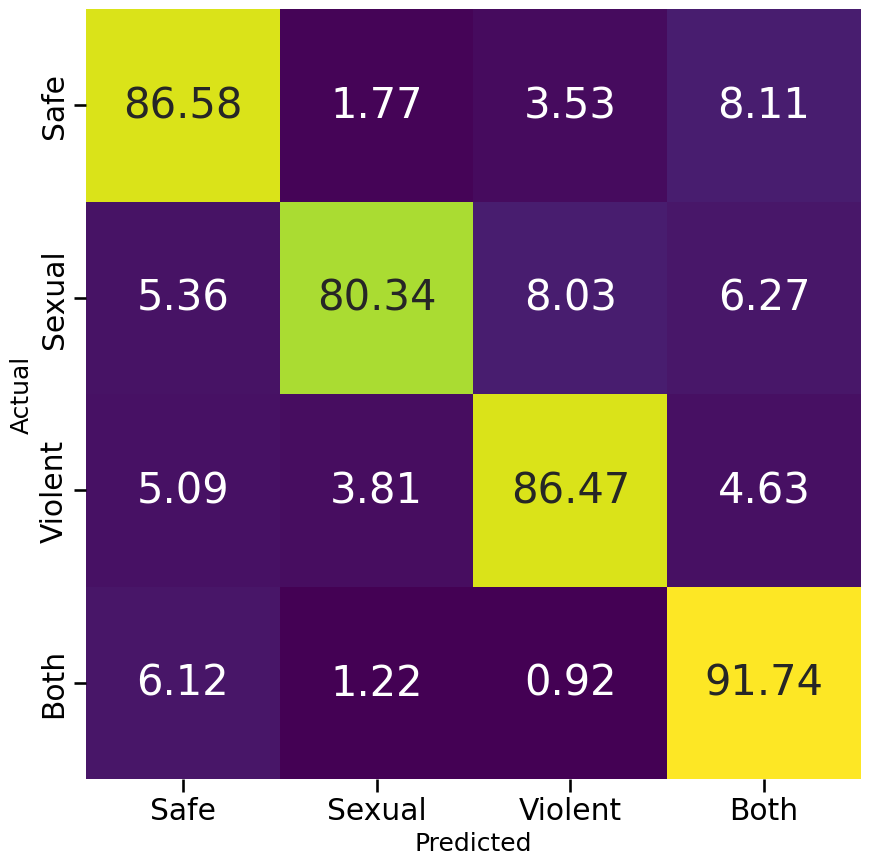}
    }
    \caption{The subfigures (a) and (b) represents the t-SNE plot visualizatoin and confusion matrix of \textbf{\texttt{SNIFR}} respectively}
    \label{tsnecm}
\end{figure}

\begin{figure}[!bt]
    \centering
    \includegraphics[width=1.0\linewidth]{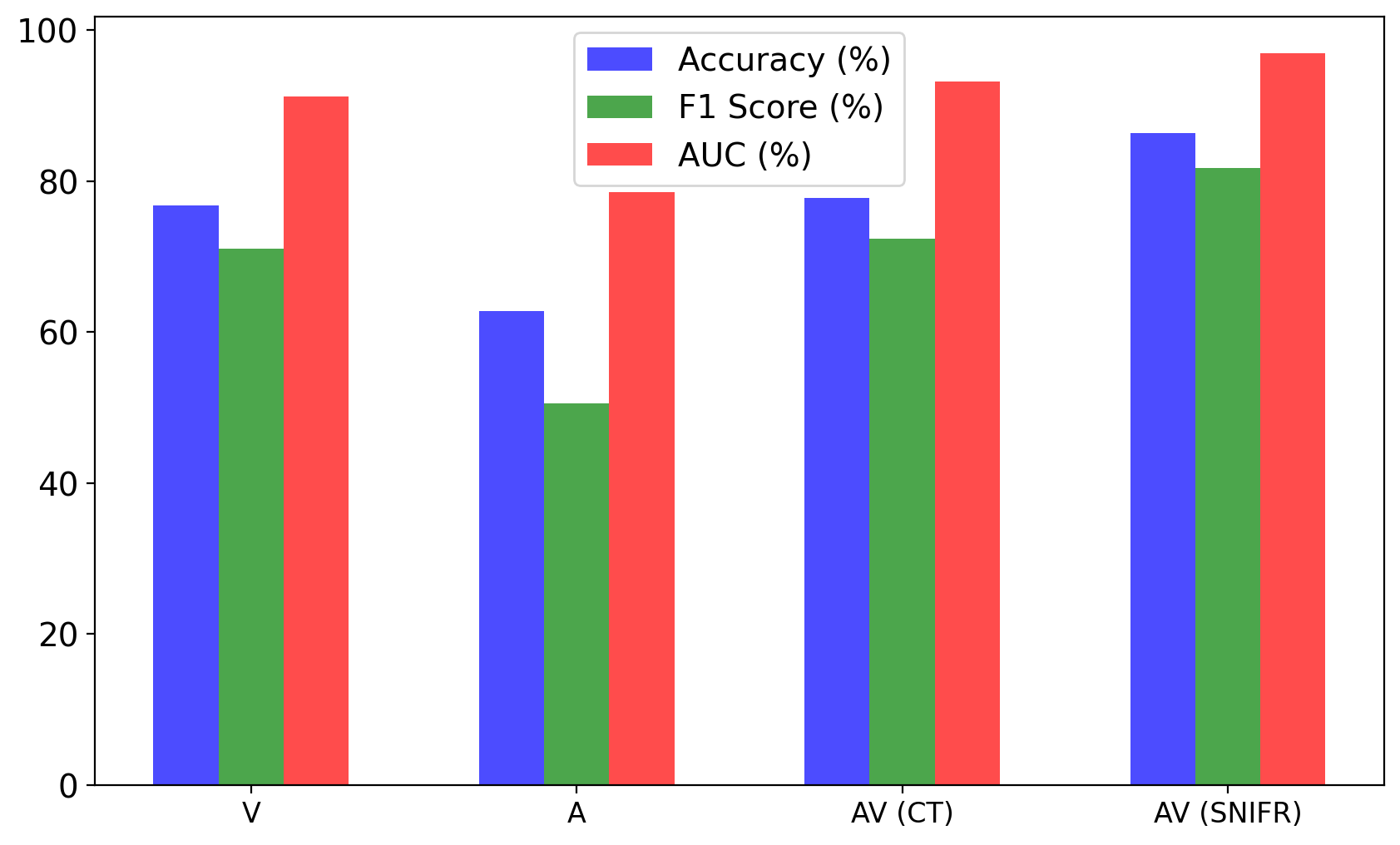}
    \caption{Total Accuracy, F1 Score, AUC; V: Visual, A: Audio, AV (CT): Cross-Transformer,  AV (SNIFR): Proposed framework, \textbf{\texttt{SNIFR}}}
    \label{Bar_plot}
\end{figure}

Table \ref{results} presents the class-wise evaluation results for models trained using single-modality and multimodal approaches for FGCHCD. Among unimodal models, visual-based models consistently outperform audio-based models, highlighting the dominance of visual cues in content classification. However, while weaker in isolation, audio models capture critical cues such as alarming sound effects, background music, and suggestive dialogue, which contribute significantly to classification. Despite this, unimodal audio models struggle with subtle harmful content when visual cues are dominant, reinforcing the necessity of multimodal fusion. However, we hypothesize that integrating audio with visual modalities will improve FGCHCD. Our experiments with the proposed novel framework, \textbf{\texttt{SNIFR}} validates our hypothesis in terms of highest accuracy, F1-score, and AUC across all the classes. We also present the total accruacy, F1-score, and AUC in Figure \ref{Bar_plot} that is obtained through average of the class-wise scores. We observe the top performance of \textbf{\texttt{SNIFR}} with audio-visual alignment. To benchmark \textbf{\texttt{SNIFR}} effectiveness, we also compare it against various baseline fusion techniques, including EC, LC, EA, and EP. The results reveal that these baseline approaches fail to effectively integrate audio-visual cues, often underperforming even compared to unimodal visual models. In contrast, \textbf{\texttt{SNIFR}} shows top performance as the structured integration of modalities in a cascaded manner through the cascaded cross-transformer blocks allows \textbf{\texttt{SNIFR}} to dynamically capture fine-grained dependencies between audio and visual cues, leading to substantial improvements. We present an ablation study of \textbf{\texttt{SNIFR}}, if instead of cascaded cross-transformers (AV (CT) in Table \ref{results}), we only use a single level of cross-transformer and that make use of cross-attention mechanism that has demonstrated effectiveness for multimodal fusion \cite{10094306, 10096732}. We see that it is not able to keep up the results attained by cascaded cross-transformers. We also plot the t-SNE plot visualization of the features from the penultimate layer of \textbf{\texttt{SNIFR}} in Figure \ref{tsnecm} (a) followed by the confusion matrix of \textbf{\texttt{SNIFR}} Figure \ref{tsnecm} (b).

%demonstrating superior accuracy in detecting explicit and violent content. Among vision-based approaches, VideoMAE captures rich spatial and temporal information, making it particularly effective for identifying subtle and transient harmful content embedded within video sequences. 

\subsection{Comparison to SOTA}

In this section, we present the comparison of the proposed framework, \textbf{\texttt{SNIFR}} with previous SOTA work. As FGCHCD is a underexplored task, so we are not able to find any other studies that worked on the same dataset, to the best of our knowledge, as used in our experiments \cite{singh2019kidsguard}. So, we compare our results with Singh et al. \cite{singh2019kidsguard} as the SOTA study where they used only-visual modality for their experiments. We observe sufficient gain in the performance across all the scores with \textbf{\texttt{SNIFR}} (See Table \ref{results}). These results mirrors our hypothesis that audio cues will act as complementary modality to visual modality for FGCHCD and also showed the superior cross-modality alignment capability of \textbf{\texttt{SNIFR}}. We acknowledge that few works have worked on similar topics, however, they haven't made their dataset public \cite{yousaf2022deep} and haven't labelled their dataset exactly for FGCHCD \cite{binh2022samba}.

% \section{Conclusion}
% \textcolor{red}{O to Swarup: Refer the abstract and frame the conclusion as you want} \newline

% \noindent In this paper, we present a novel framework for fine-grained detection of child-unsafe content in response to the increasing child audience on video-sharing platforms and the challenges posed by malicious content embedded subtly within videos. While prior studies have advanced methods for identifying explicit, violent, or sexual content, they predominantly rely on visual analysis, often overlooking valuable audio cues. Our approach addresses this gap by introducing a multimodal method that combines audio and visual signals. Experimental results demonstrate that integrating both audio and visual cues significantly enhances the detection accuracy of child-unsafe content compared to visual-only approaches.

\section{Conclusion}
In this study, we explore for the first time, aligning audio and visual modalities for FGCHCD and propose \textbf{\texttt{SNIFR}}, a novel framework for effective alignment. By leveraging a transformer encoder for intra-modality interactions and a cascaded cross-transformer for inter-modality alignment, \textbf{\texttt{SNIFR}} effectively captures dependencies between audio and visual cues, outperforming unimodal and baseline fusion models. This work highlights the crucial role of audio in enhancing visual-only models, setting new SOTA performance for FGCHCD and demonstrating the potential of multimodal integration in content moderation.

\bibliographystyle{IEEEtran}
\bibliography{main}

\begin{thebibliography}{10}
\providecommand{\url}[1]{#1}
\csname url@samestyle\endcsname
\providecommand{\newblock}{\relax}
\providecommand{\bibinfo}[2]{#2}
\providecommand{\BIBentrySTDinterwordspacing}{\spaceskip=0pt\relax}
\providecommand{\BIBentryALTinterwordstretchfactor}{4}
\providecommand{\BIBentryALTinterwordspacing}{\spaceskip=\fontdimen2\font plus
\BIBentryALTinterwordstretchfactor\fontdimen3\font minus \fontdimen4\font\relax}
\providecommand{\BIBforeignlanguage}[2]{{%
\expandafter\ifx\csname l@#1\endcsname\relax
\typeout{** WARNING: IEEEtran.bst: No hyphenation pattern has been}%
\typeout{** loaded for the language `#1'. Using the pattern for}%
\typeout{** the default language instead.}%
\else
\language=\csname l@#1\endcsname
\fi
#2}}
\providecommand{\BIBdecl}{\relax}
\BIBdecl

\bibitem{alshamrani2020hiding}
S.~Alshamrani, A.~Abusnaina, and D.~Mohaisen, ``Hiding in plain sight: A measurement and analysis of kids’ exposure to malicious urls on youtube,'' in \emph{2020 IEEE/ACM Symposium on Edge Computing (SEC)}.\hskip 1em plus 0.5em minus 0.4em\relax IEEE, 2020, pp. 321--326.

\bibitem{aldahoul2021evaluation}
N.~Aldahoul, H.~A. Karim, M.~H.~L. Abdullah, A.~S.~B. Wazir, M.~F.~A. Fauzi, M.~J.~T. Tan, S.~Mansor, and H.~S. Lyn, ``An evaluation of traditional and cnn-based feature descriptors for cartoon pornography detection,'' \emph{IEEE Access}, vol.~9, pp. 39\,910--39\,925, 2021.

\bibitem{chuttur2022multi}
M.~Y. Chuttur and A.~Nazurally, ``A multi-modal approach to detect inappropriate cartoon video contents using deep learning networks,'' \emph{Multimedia Tools and Applications}, vol.~81, no.~12, pp. 16\,881--16\,900, 2022.

\bibitem{gkolemi2022youtubers}
M.~Gkolemi, P.~Papadopoulos, E.~Markatos, and N.~Kourtellis, ``Youtubers not madeforkids: Detecting channels sharing inappropriate videos targeting children,'' in \emph{Proceedings of the 14th ACM Web Science Conference 2022}, 2022, pp. 370--381.

\bibitem{ramesh2022beach}
K.~Ramesh, A.~R. KhudaBukhsh, and S.~Kumar, ``‘beach’to ‘bitch’: Inadvertent unsafe transcription of kids’ content on youtube,'' in \emph{Proceedings of the AAAI Conference on Artificial Intelligence}, vol.~36, no.~11, 2022, pp. 12\,108--12\,118.

\bibitem{aggarwal2023protecting}
S.~Aggarwal and D.~K. Vishwakarma, ``Protecting our children from the dark corners of youtube: A cutting-edge analysis,'' in \emph{2023 4th IEEE Global Conference for Advancement in Technology (GCAT)}.\hskip 1em plus 0.5em minus 0.4em\relax IEEE, 2023, pp. 1--5.

\bibitem{alqahtani2023children}
S.~I. Alqahtani, W.~M. Yafooz, A.~Alsaeedi, L.~Syed, and R.~Alluhaibi, ``Children’s safety on youtube: A systematic review,'' \emph{Applied Sciences}, vol.~13, no.~6, p. 4044, 2023.

\bibitem{7906950}
R.~Kaushal, S.~Saha, P.~Bajaj, and P.~Kumaraguru, ``Kidstube: Detection, characterization and analysis of child unsafe content \& promoters on youtube,'' in \emph{2016 14th Annual Conference on Privacy, Security and Trust (PST)}, 2016, pp. 157--164.

\bibitem{papadamou2020disturbed}
K.~Papadamou, A.~Papasavva, S.~Zannettou, J.~Blackburn, N.~Kourtellis, I.~Leontiadis, G.~Stringhini, and M.~Sirivianos, ``Disturbed youtube for kids: Characterizing and detecting inappropriate videos targeting young children,'' in \emph{Proceedings of the international AAAI conference on web and social media}, vol.~14, 2020, pp. 522--533.

\bibitem{binh2022samba}
L.~Binh, R.~Tandon, C.~Oinar, J.~Liu, U.~Durairaj, J.~Guo, S.~Zahabizadeh, S.~Ilango, J.~Tang, F.~Morstatter \emph{et~al.}, ``Samba: Identifying inappropriate videos for young children on youtube,'' in \emph{Proceedings of the 31st ACM International Conference on Information \& Knowledge Management}, 2022, pp. 88--97.

\bibitem{balat2024tikguard}
M.~Balat, M.~Gabr, H.~Bakr, and A.~B. Zaky, ``Tikguard: A deep learning transformer-based solution for detecting unsuitable tiktok content for kids,'' in \emph{2024 6th Novel Intelligent and Leading Emerging Sciences Conference (NILES)}.\hskip 1em plus 0.5em minus 0.4em\relax IEEE, 2024, pp. 337--340.

\bibitem{singh2019kidsguard}
S.~Singh, R.~Kaushal, A.~B. Buduru, and P.~Kumaraguru, ``Kidsguard: fine grained approach for child unsafe video representation and detection,'' in \emph{Proceedings of the 34th ACM/SIGAPP symposium on applied computing}, 2019, pp. 2104--2111.

\bibitem{yousaf2022deep}
K.~Yousaf and T.~Nawaz, ``A deep learning-based approach for inappropriate content detection and classification of youtube videos,'' \emph{IEEE Access}, vol.~10, pp. 16\,283--16\,298, 2022.

\bibitem{gong21b_interspeech}
Y.~Gong, Y.-A. Chung, and J.~Glass, ``{AST: Audio Spectrogram Transformer},'' in \emph{Proc. Interspeech 2021}, 2021, pp. 571--575.

\bibitem{tong2022videomae}
Z.~Tong, Y.~Song, J.~Wang, and L.~Wang, ``Videomae: Masked autoencoders are data-efficient learners for self-supervised video pre-training,'' \emph{Advances in neural information processing systems}, vol.~35, pp. 10\,078--10\,093, 2022.

\bibitem{vaswani2017attention}
A.~Vaswani, ``Attention is all you need,'' \emph{Advances in Neural Information Processing Systems}, 2017.

\bibitem{gong2022uavm}
Y.~Gong, A.~H. Liu, A.~Rouditchenko, and J.~Glass, ``Uavm: Towards unifying audio and visual models,'' \emph{IEEE Signal Processing Letters}, vol.~29, pp. 2437--2441, 2022.

\bibitem{10094306}
J.~Lin, X.~Cai, H.~Dinkel, J.~Chen, Z.~Yan, Y.~Wang, J.~Zhang, Z.~Wu, Y.~Wang, and H.~Meng, ``Av-sepformer: Cross-attention sepformer for audio-visual target speaker extraction,'' in \emph{ICASSP 2023 - 2023 IEEE International Conference on Acoustics, Speech and Signal Processing (ICASSP)}, 2023, pp. 1--5.

\bibitem{10096732}
Z.~Xu, X.~Fan, and M.~Hasegawa-Johnson, ``Dual-path cross-modal attention for better audio-visual speech extraction,'' in \emph{ICASSP 2023 - 2023 IEEE International Conference on Acoustics, Speech and Signal Processing (ICASSP)}, 2023, pp. 1--5.

\end{thebibliography}

\end{document}